\documentclass[prd,aps,superscriptaddress,nofootinbib,amsmath,amssymb,showpacs]{revtex4}
\usepackage{graphicx}
\usepackage{dcolumn}
\usepackage{bm}
\usepackage{multirow}
\usepackage{adjustbox}
\usepackage{textcomp}
\usepackage{mathtext}
\usepackage{hyperref}
\graphicspath{{./eps_KDAR/}}

\begin{document}

\title{\Large The low energy inclusive $\nu_\mu(\nu_e)$ - $^{12}$C scattering revisited}
\author{M. Sajjad \surname{Athar}}
\email{sajathar@gmail.com}
\affiliation{Department of Physics, Aligarh Muslim University, Aligarh-202002, India}
\author{S. K. \surname{Singh}}
\affiliation{Department of Physics, Aligarh Muslim University, Aligarh-202002, India}
\begin{abstract}
We have reviewed the current status of the inclusive  neutrino scattering from $^{12}$C in the low energy region corresponding 
to the neutrino beams from the pion, muon and kaon decaying at rest. The theoretical calculations of total cross sections in 
various nuclear models with special emphasis on the recent experiments with the monoenergetic neutrinos from KDAR~\cite{Aguilar-Arevalo:2018ylq} along with the 
older experiments from KARMEN and LSND collaborations have been discussed in the context of the recent works by Akbar et 
al.~\cite{Akbar:2017dih} and Nikolakopoulos et al.~\cite{Nikolakopoulos:2020alk}. The inadequacy of the various theoretical models used to explain the 
experimental results on the inclusive neutrino scattering from nuclei at low energies has been highlighted and the need for a 
better understanding of the nuclear medium effects beyond the impulse approximation has been emphasized.
 \end{abstract}
 \pacs{{12.15.-y}, {13.15.+g}} 
\maketitle
 \section{Introduction}
 The inclusive neutrino scattering in $^{12}$C at low energies has been studied as a background of the neutrino oscillation
 experiments performed with the liquid scintillator detectors which contain $^{12}$C as the target. The first experiments in the 
 low energy region were done almost 40 years ago at the Brookhaven National Laboratory with the muon neutrinos~($\nu_{\mu}$) 
 as well as with the electron neutrinos~($\nu_{e}$)~\cite{Cortex}. Since then quite a few experiments have measured the inclusive 
 cross sections with $\nu_{\mu}$ and $\nu_e$ in $^{12}$C nucleus in the low energy region i.e. $E_{\nu_{\mu}} \le 280$~MeV and 
 $E_{\nu_{e}} \le 52.8$~MeV~\cite{Koetke:1992yk,Albert:1994xs,Athanassopoulos:1997rn,Auerbach:2002iy,Krakauer:1991rf,
 Bodmann:1994py,Kretschmer:2002iq,Maschuw:1998qh,Athanassopoulos:1997rm,Auerbach:2001hz}. Theoretical calculations of these cross 
 sections in $^{12}$C were done quite early by many authors using 
 various versions of the shell model and the Fermi gas model to describe the structure of the $^{12}$C nucleus which are summarized 
 in excellent reviews of Llewellyn Smith~\cite{LlewellynSmith:1971uhs}, Walecka~\cite{Walecka}, and Donnelly and 
 Peccei~\cite{Donnelly:1978tz}. However, the recent topical interest in studying the inclusive neutrino scattering in $^{12}$C at low 
 energies was regenerated again when the experimental results from the LSND collaboration at LANL were found inconsistent with the 
 existing theoretical predictions specially in the case of $\nu_{\mu}$ scattering~\cite{Koetke:1992yk,Albert:1994xs,
 Athanassopoulos:1997rn,Auerbach:2002iy}. This led to many new theoretical calculations for the neutrino scattering in $^{12}$C 
 which were done using various nuclear models, for example the various versions of 
 sophisticated microscopic models~\cite{Nikolakopoulos:2020alk,Kolbe:1994xb,Kolbe:1995af,Kolbe:1999au,Kolbe:2003ys,Volpe:2000zn,Auerbach:1997ay,
 Hayes:1999ew,Auerbach:2002tw,Paar:2008zza,Krmpotic:2004gx,Pandey:2014tza,Jachowicz:2002rr,Gonzalez-Jimenez:2019qhq} and the 
 relativistic Fermi gas models~\cite{Smith:1972xh,Singh:1993rg,Umino:1994wu,Umino:1996cz,Gaisser:1986bv,Kosmas:1996fh,
 Singh:1998md,Nieves:2004wx,SajjadAthar:2005ke,Nieves:2017lij,Valverde:2006zn,Vagnoni:2017hll,Akbar:2017dih,Ivanov:2018nlm} as well as some other 
 models~\cite{Mintz:1995ww,Kim:1985zs,Frazier:1970rb,Kim:1979,Kubodera:1993rk}, to describe the structure of $^{12}$C. Most of these calculations were done in the 
 impulse approximation(IA) and emphasized the need to go beyond the impulse approximation by including the effects due to the 
 nucleon-nucleon correlations and meson exchange currents~(MEC) in the nuclear medium. However, not much work was done to study 
 these effects except the work of Hayes and Towner~\cite{Hayes:1999ew} and Umino et 
 al.~\cite{Umino:1994wu,Umino:1996cz} in the context of the LSND experiment.
  
 In recent times many experiments have measured the inclusive neutrino scattering cross sections in $^{12}$C with $\nu_{\mu}$ in 
 the intermediate energy range of $E_{\nu_{\mu}} = 0.7- 1.2$~GeV and a theoretical understanding of these experiments also 
 emphasizes the need for including the effects due to the nucleon-nucleon correlations and meson exchange 
 currents~\cite{Alvarez-Ruso:2017oui,Alvarez-Ruso:2014bla,Morfin:2012kn,Formaggio:2013kya,Gallagher:2011zza,Katori:2016yel,
 Benhar:2013bwa,Benhar:2015wva}. The muon neutrinos, in these experiments, are produced by the decay of pions and 
 kaons  (created in the scattering of the protons from nuclei in the proton accelerators) in flight and arrive at the target-cum-detector 
 with a continuous energy spectrum having a peak around 0.6GeV. A precise determination of the energy spectrum 
 of $\nu_\mu$ used in these experiments is made difficult due to the uncertainties in the knowledge of the hadronic interactions 
 and the dynamics of proton-nucleus scattering involved in the production of pions and kaons in the energy region of a few GeV of 
 protons. In most of these neutrino experiments, the energy spectrum of the incident $\nu_{\mu}$ is reconstructed using the 
 dynamics of the two body quasielastic~(QE) neutrino reactions in nuclei. This QE dynamics is affected by the various nuclear 
 medium effects like the binding energy and the Fermi motion of the nucleons as well as the nucleon-nucleon correlations which are 
 needed to be taken into account. The corrections due to these nuclear medium effects in the reconstruction of incident neutrino 
 spectrum are incorporated but they are model dependent and lead to considerable uncertainties in determining the neutrino 
 spectrum and the flux averaged inclusive neutrino nucleus cross sections~\cite{Alvarez-Ruso:2017oui,Alvarez-Ruso:2014bla,
 Morfin:2012kn,Katori:2016yel,Benhar:2015wva}. In this context, attempts have been recently made to develop the monoenergetic neutrino beams
 at the Fermilab~\cite{Spitz:2012gp,Spitz:2014hwa,Axani:2015dha,
 Acciarri:2016smi,Antonello:2015lea,Fava:2019fuz} and JPARC~\cite{Ajimura:2017fld,Harada:2013yaa,Park:2020uck,Park:2020vxw} laboratories 
 specially in the low energy region using kaon decaying at rest~(KDAR). The kaons decay through the $Kl_2$, $Kl_3$ and $Kl_4$ 
 decay modes ($l=e,~\mu$). 
  The muon neutrinos from $K^{+} \rightarrow \mu^{+} + \nu_{\mu}$ decay are monoenergetic 
  with an energy $E_{\nu_\mu} = 236$~MeV accompanied by $\nu_{e}$ and $\nu_{\mu}$ 
 from $K_{e 3}$ and $K_{\mu 3}$ decays, respectively, with continuous energy spectra. The MiniBooNE collaboration at the 
 Fermilab~\cite{Aguilar-Arevalo:2018ylq} has recently reported the first measurement of the inclusive muon neutrino cross sections 
 with monoenergetic neutrinos of $E_{\nu_{\mu}} = 236$~MeV. A comparison of the experimental result with the recent theoretical 
 calculations made for the monoenergetic KDAR neutrinos~\cite{Akbar:2017dih,Nikolakopoulos:2020alk}  shows a clear 
 discrepancy between the theoretical and the experimental results. This discrepancy is similar to the discrepancy encountered 
 almost 20 years ago in the case of LSND neutrino experiment using muon neutrinos with a continuous energy spectrum~\cite{Koetke:1992yk,Albert:1994xs,Athanassopoulos:1997rn,Auerbach:2002iy} in the low energy region of $E_{\nu_{\mu}} 
 \le 280$~MeV. This has motivated us to revisit the subject of the inclusive neutrino scattering in $^{12}$C at low energies in 
 the context of the new experiment with the monoenergetic KDAR $\nu_\mu$ and emphasize the role of the nuclear medium 
 effects~(NME). 
 
 In sections~\ref{sec2}A and ~\ref{sec2}B, we present a review of the experimental and theoretical works in the low energy inclusive 
 neutrino scattering in $^{12}$C in the 
 context of KARMEN and LSND experiments and discuss in section~\ref{sec3}, the results of the recent MiniBooNE experiment in the intermediate 
 energy range of 0.6-1.2GeV and the latest theoretical developments in understanding the nuclear medium effects 
 in the QE neutrino-nucleus scattering. 
 In section~\ref{sec4}, we have discussed the experimental as well as theoretical results of inclusive cross sections with 
 monoenergetic KDAR neutrinos with $E_{\nu_\mu}$=236MeV in the impulse approximation as well as some effects beyond the impulse approximation.
 We conclude in section~\ref{Summary} with some comments emphasizing the need for further theoretical work to explain the earlier 
 and recent results in the low energy inclusive neutrino scattering in $^{12}$C 
 which could help to analyze the results expected from the future experiments planned at the Fermilab and at the 
 JPARC in $^{12}$C and $^{40}$Ar nuclear targets.
 
 \section{Inclusive weak nuclear processes at low energies}\label{sec2}
 \subsection{Experimental results}
  In the low energy region, the following weak inclusive processes have been studied in $^{12}$C using $\mu^{-}$, $\nu_{\mu}$ and 
  $\nu_{e}$ beams:
  \begin{eqnarray}
   \mu^{-} + ^{12}C &\longrightarrow& \nu_{\mu} + X, \\
   \nu_{\mu} + ^{12}C &\longrightarrow& \mu^{-} + X, \\
   \nu_{e} + ^{12}C &\longrightarrow& e^{-} + X.
  \end{eqnarray}
The study of the low energy weak inclusive processes in nuclei started with the attempts made to understand the early experimental 
results in the inclusive nuclear muon capture in various nuclei including $^{12}$C. Accordingly many microscopic calculations were 
done for the inclusive muon capture in $^{12}$C employing quite elaborate nuclear models to obtain the nuclear wave functions of 
the initial and final nuclei. The transition rates to the ground state as well as to the various excited states in the final 
nucleus were calculated and summed over to obtain the inclusive muon capture rate. In addition to these explicit microscopic 
calculations, studies were also made by applying the closure approximation over the final nuclear states using 
many particle shell model wave functions of the initial state. Furthermore, similar calculations in various verminous versions of the relativistic Fermi gas 
model and the other models were also done. A comparison 
of the theoretical results on inclusive muon capture with the available experimental results emphasized the need for understanding 
various nuclear medium effects beyond the impulse approximation like the effects due to the nucleon-nucleon correlations, residual 
interaction and pairing between the nucleons as well as the meson exchange currents in nuclei. The earlier works in the inclusive muon capture in nuclei using these methods are summarized in excellent reviews by 
Mukhopadhyay~\cite{Mukhopadhyay:1976hu}, Measday~\cite{Measday:2001yr},  Auerbach and 
Klein~\cite{Auerbach:1984ph} and Suzuki et al.~\cite{Suzuki:1987jf}.
 These effects were also found to be 
quite important in the other low energy weak process like $p + p \rightarrow d + e^+ + \nu_e$ and leading to the renormalization of the axial 
coupling in the nuclear beta decays~\cite{Dautry:1975xq,Ohta:1974fa,Ericson:1973vj,Riska:1970jxh, Nieves:2011yp,Delorme:1985ps,Muller:1981dc}. 

As mentioned in the introduction, the importance of these nuclear medium effects was reiterated again when the inclusive neutrino 
scattering cross sections at low energies were measured by the KARMEN experiment at the ISIS facility at the  Rutherford Appleton 
Laboratory using $\nu_{e}$ beams~\cite{Bodmann:1994py,Kretschmer:2002iq,Maschuw:1998qh} and by the LSND experiment at the 
LAMPF~\cite{Krakauer:1991rf,Athanassopoulos:1997rn} with $\nu_{e}$ and $\nu_{\mu}$ beams. 

\begin{figure}
\includegraphics[height=5cm,width=8cm]{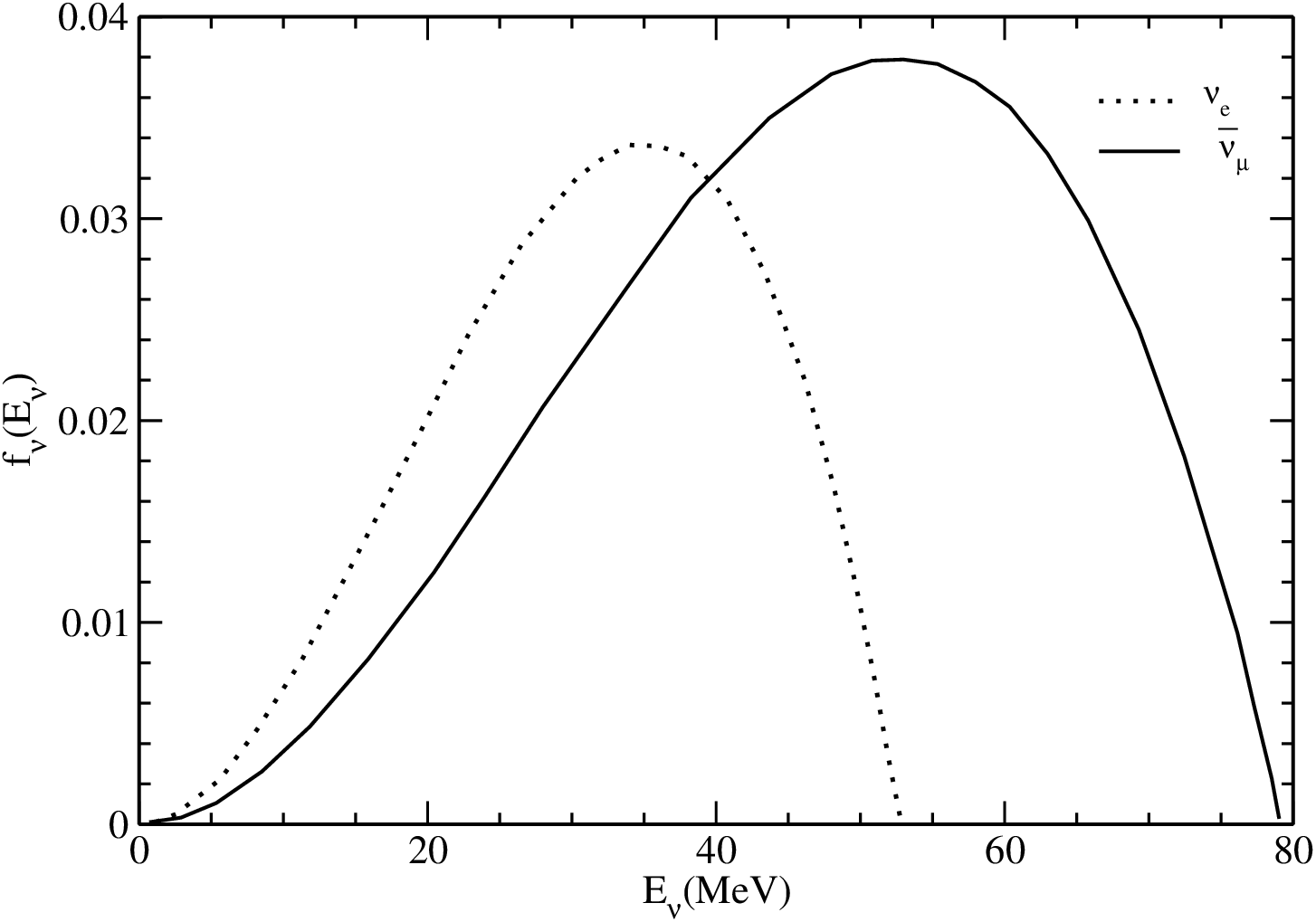}
\caption{Michel spectra, showing the $\nu_e$ (dotted line) and $\bar\nu_\mu$ (solid line) distributions originating from muon 
decay at rest ($\mu$-DAR).}\label{Michel-spectra}
\end{figure}

In the case of $\nu_{e}$ beams, the inclusive cross sections were measured in both experiments and the results are reported in many 
publications~\cite{Krakauer:1991rf,Bodmann:1994py,Kretschmer:2002iq, Maschuw:1998qh,Athanassopoulos:1997rm}. The $\nu_{e}$ and 
$\bar\nu_{\mu}$ beams are obtained from the muons decaying at rest~($\mu$DAR) with $E_{\nu_{e}} \le 52.8$~MeV and the neutrino 
energy spectrum described by the Michel spectrum, for $\nu_e$ is given by
\begin{eqnarray}
 f(E_{\nu_e}) &=& \frac{96 {E_\nu}^2}{{m_\mu}^4} (m_\mu - 2 E_{\nu}),
\end{eqnarray}
and for $\bar\nu_{\mu}$ by
\begin{eqnarray}
 f(E_{\bar \nu_\mu}) &=& \frac{16 {E_\nu}^2}{{m_\mu}^4} (3 m_\mu - 4 E_{\nu}),
\end{eqnarray}
as shown in Fig.\ref{Michel-spectra}~\cite{Divari:2012cj}. It should be emphasized here that the $\nu_{e}$ spectrum 
resembles, to a great extent, with the various simulated spectra for supernova neutrinos available in 
literature~\cite{Totani:1997vj,Duan:2010bf,Gava:2009pj}. This makes the experimental information about the inclusive cross sections 
with the $\mu$DAR neutrinos in this energy range very useful for studying the $\nu_{e}$ induced nuclear reactions in $^{12}$C and 
other nuclei for the detection of supernova neutrinos~\cite{Scholberg:2012id}.

\begin{figure}
\includegraphics[height=5cm,width=8cm]{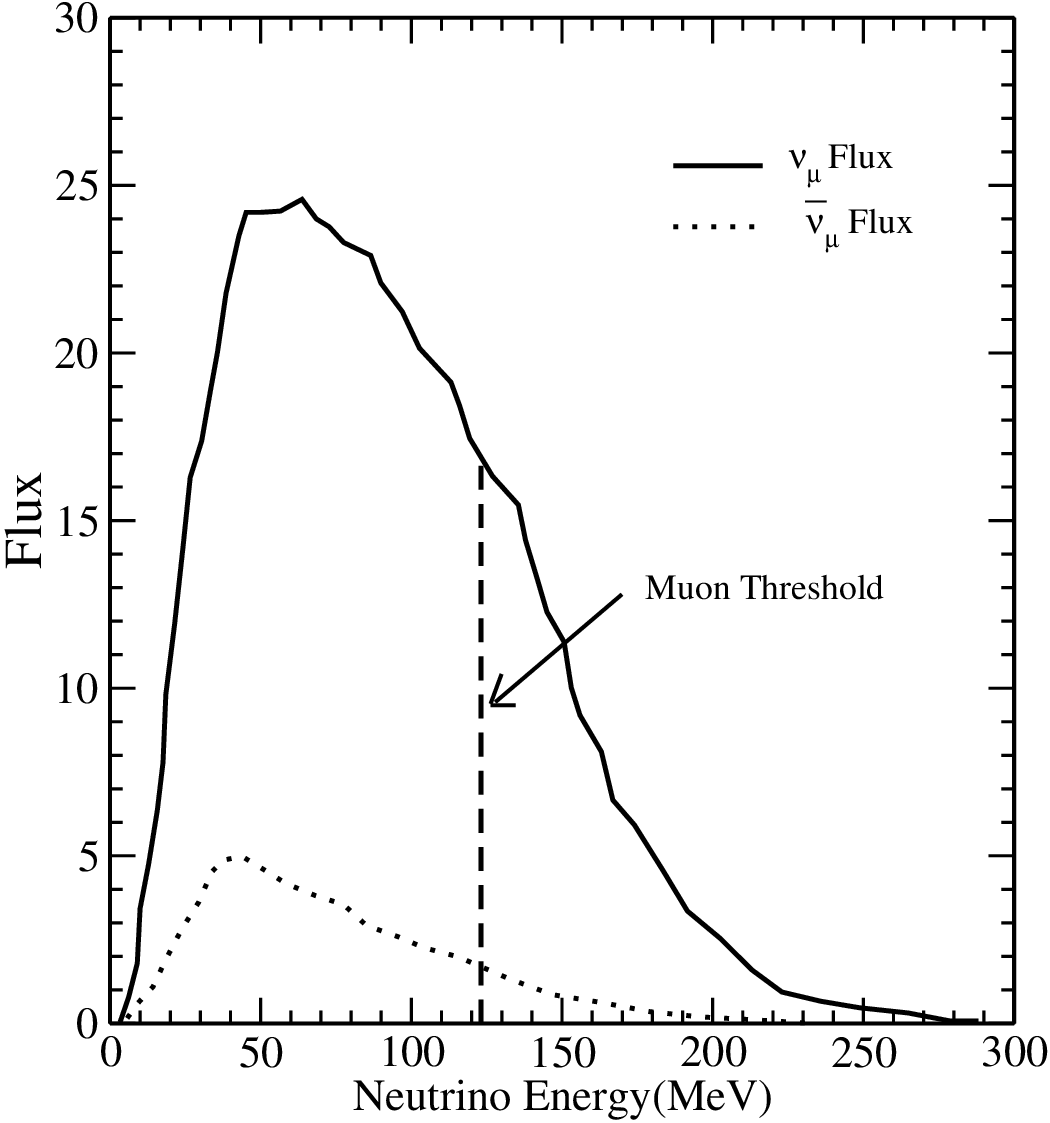}
\caption{The solid line shows $\nu_\mu$ flux from $\pi^+$ decay in flight and the dotted line shows the flux of $\bar\nu_\mu$ 
 from $\pi^-$ decay in flight for the same integrated proton beam.}
\label{DIF-spectra}
\end{figure}
In the case of $\nu_\mu$, the inclusive reaction cross sections  have been measured by the LSND collaboration at LAMPF during the 
years 1992--2002 and the results are reported in many publications~\cite{Koetke:1992yk,Albert:1994xs,Athanassopoulos:1997rn,
Auerbach:2002iy}. These experiments were done with $\nu_\mu$ coming from pions decaying in flight ($\pi$DIF) 
in which the pions are produced by the Los Alamos Neutrino Scattering Center~(LANSCE) at LAMPF using proton beam of energy $E_P 
\approx$  800 MeV. The muon neutrino spectrum is generated by a Monte-Carlo simulation which reproduces the exact 
spectrum of neutrinos from pions and muons decay at rest (DAR) quite satisfactorily. This Monte Carlo simulation program was then 
used to simulate the $\nu_\mu$ flux from the pions decay in flight. The $\nu_\mu$ fluxes for $\pi$DIF neutrinos have an 
uncertainty of about $15\%$ due to the uncertainty in the knowledge of the pion production cross section in this energy region. 
The simulated flux of the muon neutrinos from pions decay is shown in Fig.\ref{DIF-spectra}, where the vertical line shows the 
threshold energy for the muon production. The average neutrino energy above the muon production threshold ($E_{\text{th}}=123.1$ 
MeV) is around $E_\nu$=170MeV.   
  
Very recently, the inclusive $\nu_\mu$ cross section in $^{12}$C using monoenergetic KDAR neutrinos with $E_{\nu_{\mu}}$=236 
MeV has also been measured by the MiniBooNE collaboration~\cite{Aguilar-Arevalo:2018ylq}. In Table-1, we 
present all the experimental results of the low energy inclusive neutrino cross sections for $\nu_\mu$ and $\nu_e$ scattering 
from $^{12}$C reported in the literature along with the respective energy of the neutrinos. It can be seen from Table-1 that all 
the experimental results from KARMEN~\cite{Bodmann:1994py,Kretschmer:2002iq,Maschuw:1998qh} and LSND~\cite{Albert:1994xs,
Athanassopoulos:1997rn,Auerbach:2002iy} collaborations for $\nu_e$ are compatible with each other. In the case of $\nu_\mu$, all the 
experimental results are also compatible with each other except for an earlier result reported from Koetke et 
al.~\cite{Koetke:1992yk} which reportedly had a different $\nu_\mu$ flux. In Table-1, we also show the recent 
results for the inclusive neutrino cross sections in the case of the monoenergetic $\nu_\mu$ from KDAR(row-3).
   
The other low energy inclusive weak process is the ordinary muon capture in $^{12}$C, which has been experimentally studied for a 
long time and the total capture rate is reported from almost a dozen experiments with varying precision~\cite{Suzuki:1987jf}. We 
quote in Table-1, some of the most referred experimental results reported by Martino~\cite{Martino:1986gq}, Eckhause et 
al.~\cite{Eckhause:1966}, Suzuki et al.~\cite{Suzuki:1987jf}, and Ishida et al.~\cite{Ishida:1986jz}.
   
\subsection{Theoretical results and comparison with experiments}\label{sec2b}
The weak processes of inclusive  muon  capture and  neutrino scattering from nuclei at low energies have been theoretically 
studied for a long time. The various theoretical calculations for these processes are done using:
\begin{itemize}
\item  [(i)]  the microscopic nuclear models  to describe the  nuclear structure of the initial and final nuclei and to calculate 
the transition rate leading to the ground state and excited state of the final nucleus. These transition rates are then summed 
over all the states to obtain the inclusive cross sections. Subsequently, these nuclear models were improved by enlarging the basis 
space in the shell model(SM) calculations and incorporating the nucleon-nucleon correlations and pairing. 
These microscopic nuclear models use various parametric forms of the nucleon-nucleon potential to calculate the nuclear states in 
the initial and final nuclei. These nuclear states are then used to calculate the transitions to the ground state as well as to 
the various excited states in the final nucleus using multipole expansion of the electroweak transition operator~\cite{Walecka, 
Donnelly:1973enn, Donnelly:1978tz}. 
 In the case  of inclusive reactions, these methods 
have limited applicability where the transitions  to only a small number of the low lying excited states in the final nucleus can 
be calculated in a  consistent and reliable way. As the initial neutrino energy($E_\nu$) increases, the number of excited states 
become large and  reliable calculation of the wave function of the higher excited states and the transition rates  to all of these 
excited states  in the final nucleus  becomes very difficult and  intractable and some approximate methods described below have 
been used. 
 
\item [(ii)]  non-relativistic global and local Fermi gas models and their relativistic versions have been used to describe the nucleus. In some of these 
 models, the 
experimentally determined spectral function of the nucleons is used to describe their momentum distribution in the nucleus. These 
theoretical models were later improved by incorporating  the long-range nucleon-nucleon correlation effect using ladder  
approximation in the calculation of particle-hole excitations~\cite{Smith:1972xh,Gaisser:1986bv,Singh:1993rg,Umino:1994wu,Umino:1996cz,
Kosmas:1996fh,Singh:1998md,Nieves:2004wx,Nieves:2017lij,Valverde:2006zn,Vagnoni:2017hll,Akbar:2017dih,Ivanov:2018nlm}.

\item [(iii)]   other nuclear models using the closure approximation for summing over the contributions from various final states 
in the case of medium and heavy nuclei or using the elementary particle models specially  in case of some light nuclei like 
$^{3}He,^{6}Li$ and $^{12}$C etc.~\cite{Mintz:1995ww,Kim:1985zs,Frazier:1970rb,Kim:1979,Kubodera:1993rk} were also used.
\end{itemize}
   
Most of the theoretical calculations mentioned above were done within the impulse approximation and no serious attempts were made 
to include the nuclear medium effects(NME) beyond the impulse approximation. Recently the importance of including these NME 
effects beyond the impulse approximation in the inclusive neutrino scattering in $^{12}C$ was highlighted again when the 
discrepancy between the experimental results and the theoretical predictions in the inclusive neutrino cross sections in the 
intermediate energy region of 500 MeV to 1 GeV were reported~\cite{AguilarArevalo:2010zc, Katori:2009du} which led to the axial 
dipole mass($M_A$) anomaly and 
attempts were made to understand this anomaly~\cite{Martini:2009uj, Martini:2010ex,Martini:2011wp,Nieves:2011pp,Nieves:2012yz}. 

Before we discuss the nuclear medium effects in the context of understanding $M_A$ anomaly in the next section, we summarize 
in this section, the representative or all the experimental results for the low energy inclusive $(\nu_e,e^-),(\nu_\mu,\mu^-)$ scattering cross 
sections from the KARMEN, LSND  and MiniBooNE collaborations and the inclusive muon capture rates in $^{12}$C, presented in 
Table-2, along with the theoretical predictions for these processes in various nuclear models. We have included theoretical 
results only from those calculations which quote results for the relevant physical observable for all the three processes i.e. 
the total cross section for the  $(\nu_e,e^-)$ and $(\nu_\mu,\mu^-)$ inclusive scattering as well as the total rates for muon 
capture in $^{12}$C using the same nuclear model.  We show in Table-2, the physical processes (column-1), the most recent experimental 
results (column-2) and the theoretical predictions for the inclusive cross sections/capture rates in the various microscopic 
nuclear models (column 3) and the Fermi gas models (column 4).

\begin{figure}
\includegraphics[height=6cm,width=8cm]{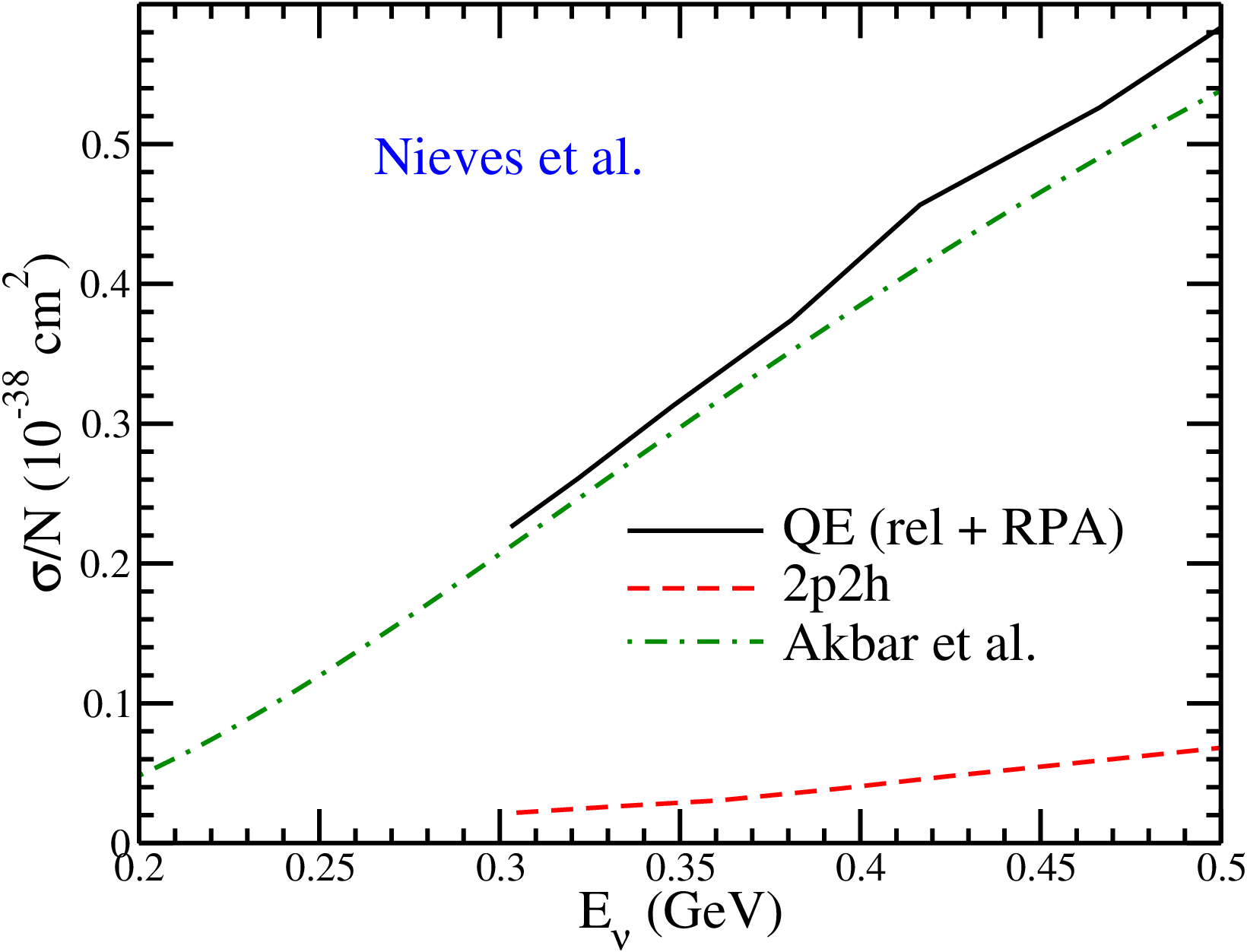}
\includegraphics[height=6cm,width=8cm]{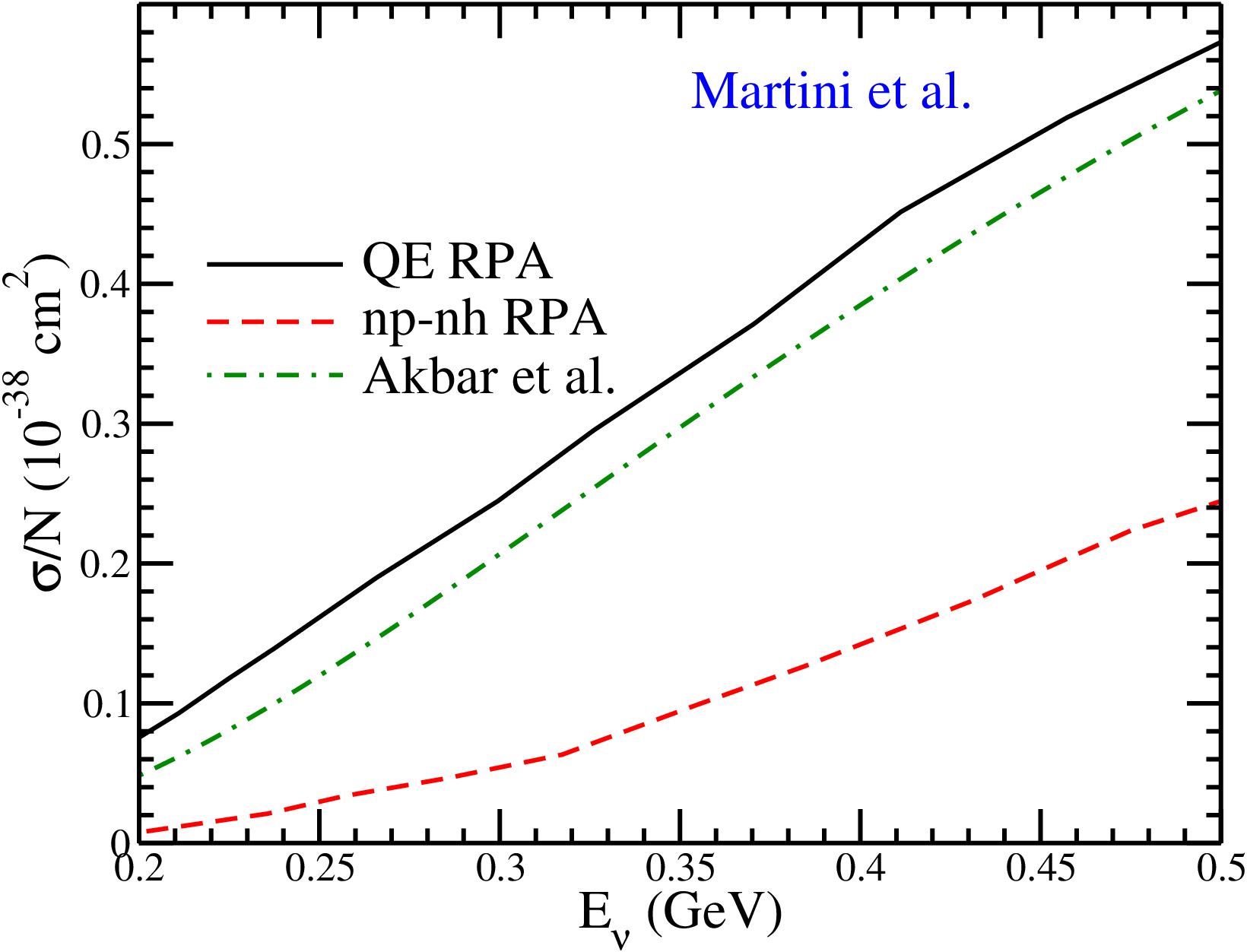}
\includegraphics[height=6cm,width=8cm]{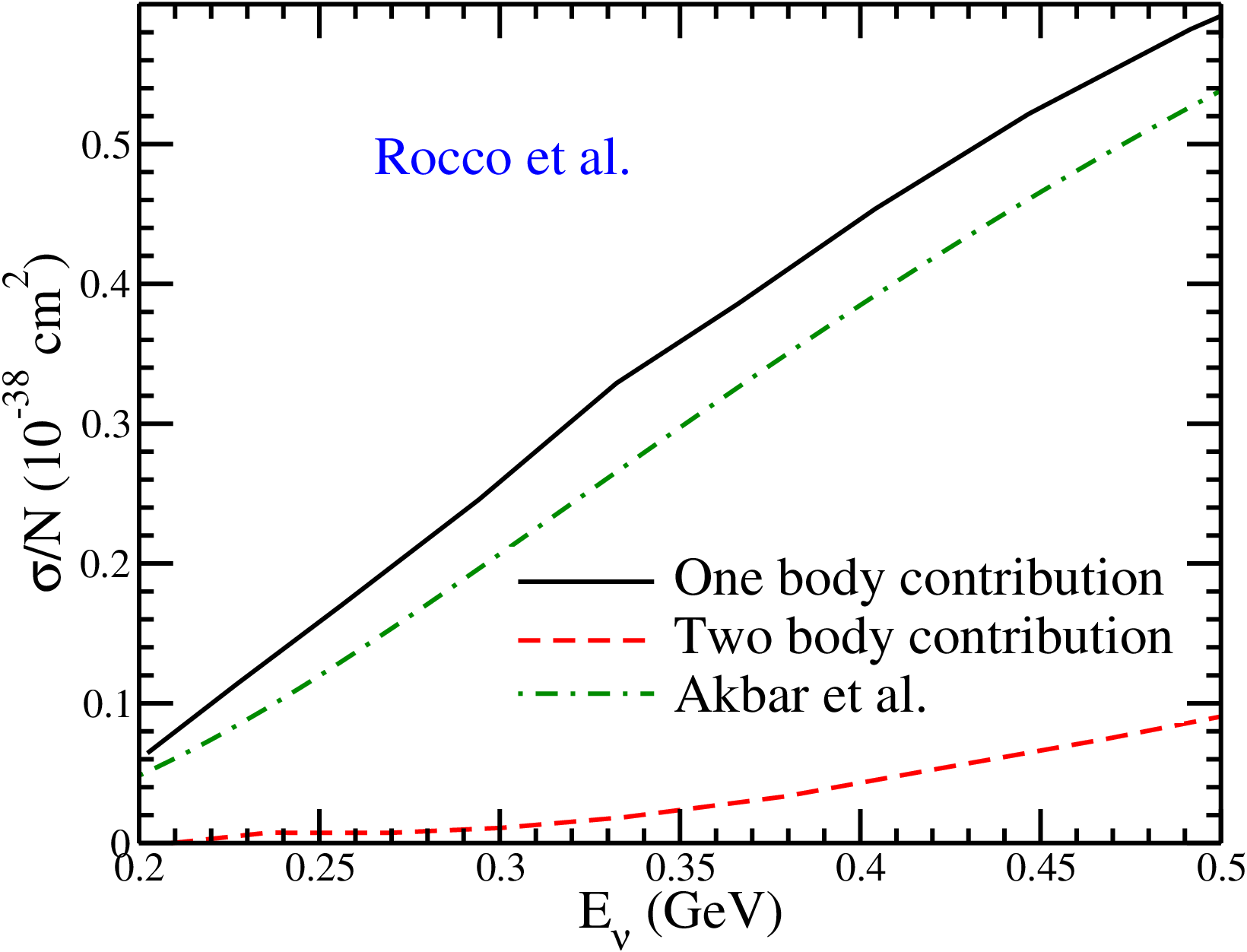}
\includegraphics[height=6cm,width=8cm]{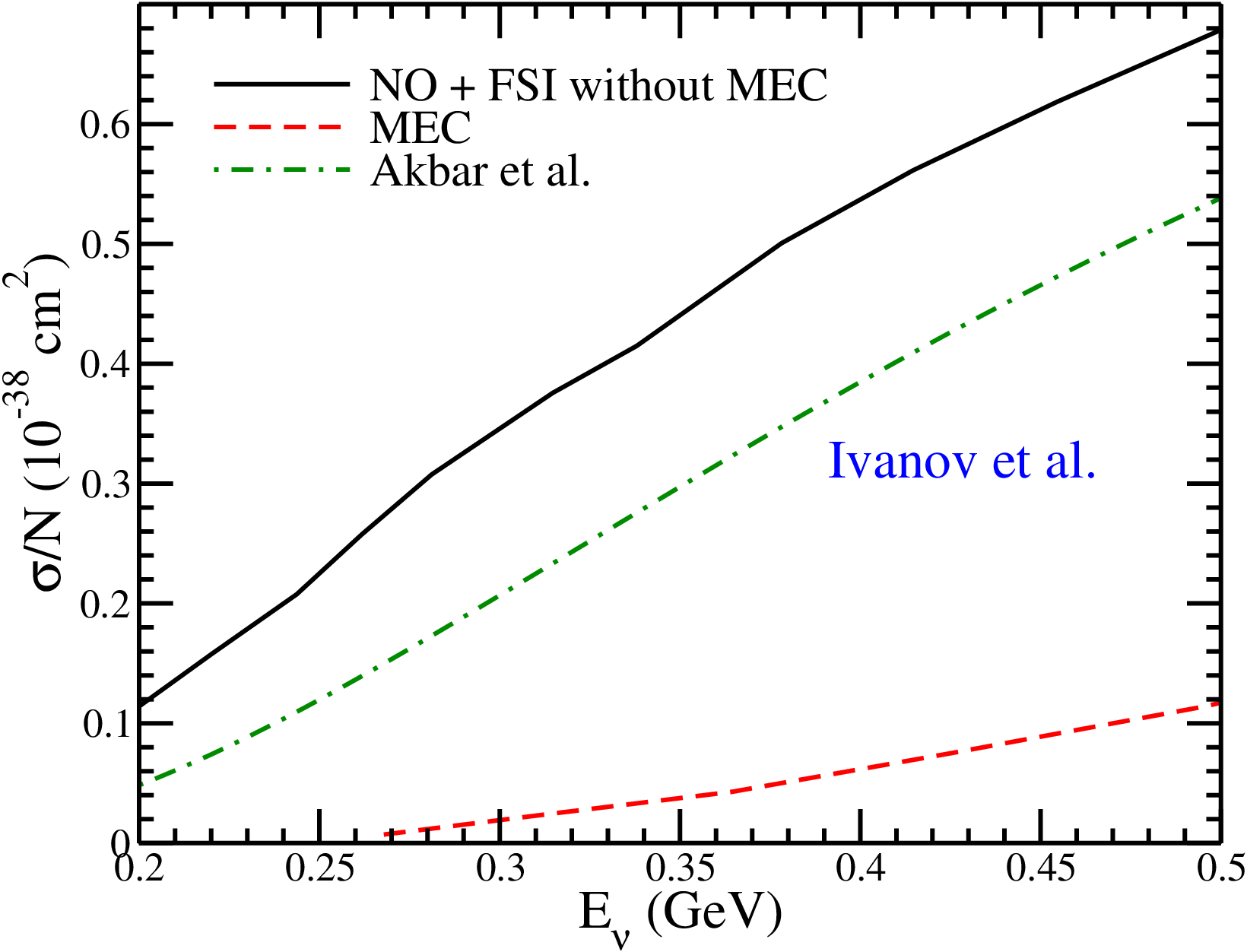}
\caption{$\nu_\mu$ scattering cross section in $^{12}C$ per neutron target. Clockwise from the top left: (1) Nieves et 
al.~\cite{Nieves:2011pp} in the relativistic Fermi gas model with RPA effect and without 2p-2h contribution(solid line) and only 
the 2p-2h contribution(dashed line); (2) Martini et al.~\cite{Martini:2010ex,Martini:2011wp} relativistic Fermi gas model with RPA 
effect and without np-nh contribution (solid line) and only the np-nh contribution(dashed line); (3) Ivanov et 
al.~\cite{Ivanov:2018nlm} using realistic spectral function with nucleon-nucleon correlations and without MEC contribution 
(solid line) and only the MEC contribution (dashed line), and (4)  Rocco et al.~\cite{Rocco:2018mwt} using impulse approximation 
with spectral function and without 2p-2h contribution, and only the 2p-2h contribution(dashed line). The results of Akbar et 
al.~\cite{Akbar:2017dih} has been shown by dash-dotted line using the relativistic Fermi gas model with RPA effect.}
\end{figure}

The microscopic nuclear models referred in the column 3 of Table-2, in general use the shell model  with large varying model 
spaces including states from 1 $\hbar \omega$ to 4 $\hbar \omega$   excitations to calculate the ground state of the initial and 
the final nuclei and transitions between them.  The various models use different forms of the phenomenological nucleon-nucleon 
potential like the Bonn potential, Landau-Migdal potential or the Skymre model potential to calculate the nuclear states. 
Moreover, the residual interaction between the nucleons which leads to the pairing and also to the quasi-particle  excitations of 
nucleons in the nuclei are also included in some nuclear models to calculate 
the nuclear wave functions of the higher excited states and the continuum state in the final nucleus. 
Depending upon the various assumptions and approximations, many nuclear models have been developed to calculate the nuclear 
wave functions of the initial and final nuclei and various transitions between them. For example, the models using RPA and CRPA by 
Kolbe et al.~\cite{Kolbe:1994xb,Kolbe:1995af,Kolbe:1999au} and Jachowicz et al.~\cite{Jachowicz:2002rr}, RPA with Hartree-Fock 
wave functions by Auerbach et al.~\cite{Auerbach:1997ay}, QRPA by Volpe et al.~\cite{Volpe:2000zn}, PQRPA by Paar et 
al.~\cite{Paar:2008zza} and RQPRPA by Krmpotic et al.~\cite{Krmpotic:2004gx} as well as the large basis elaborate multiparticle 
shell model by Hayes and Towner~\cite{Hayes:1999ew} and Auerbach and Brown~\cite{Auerbach:2002tw} have  been used to 
calculate the weak reactions given in Eq.~(1) in $^{12}$C  nucleus. The results for the inclusive cross sections from these 
calculations are shown in Table-2, column-3.

In the case of Fermi gas models, the non-relativistic global Fermi gas models were historically the first ones to be used to 
calculate these reactions~\cite{LlewellynSmith:1971uhs}. Later various versions like the relativistic Fermi gas 
model(RFG)~\cite{Smith:1972xh, Gaisser:1986bv}, non-relativistic and relativistic version of the local Fermi gas models 
(LFG)~\cite{Gaisser:1986bv,Singh:1993rg,Umino:1994wu,Umino:1996cz,Kosmas:1996fh,Singh:1998md,Nieves:2004wx,Valverde:2006zn,Nieves:2017lij,Vagnoni:2017hll} have been used to calculate these reactions in $^{12}$C. Further improvements in 
the local Fermi gas models have been incorporated by taking into account the long range nucleon-nucleon correlation using the Random Phase 
Approximation (RPA) as well as the spectral function (SF) of the nucleons in the nucleus to describe the momentum distribution of 
the nucleon to calculate the total cross section in the low energy inclusive scattering of neutrinos and inclusive muon capture in 
$^{12}$C~\cite{Nieves:2017lij,Vagnoni:2017hll, Ivanov:2018nlm}. The results from these models are shown in column-4 of Table-2. 
 
The following general observations about the theoretical results and their comparisons with the experimental results shown here in 
Table-2, can be made:
\begin{itemize}
\item [(i)] None of the microscopic nuclear model calculations is able to explain all the three weak processes in a given version of the model. For example, the 
models of Kolbe et al.~\cite{Kolbe:1994xb,Kolbe:1995af,Kolbe:1999au} using RPA and CRPA explain the $(\nu_e,e^-)$ reactions but 
overestimate the $ (\nu_\mu,\mu^-)$ and underestimate the $ (\mu^-,\nu_\mu)$  reactions. The RPA and QRPA models of Volpe et 
al.~\cite{Volpe:2000zn} overestimates all the processes and PQPRPA  model of Krmpotic et al.~\cite{Krmpotic:2004gx} explains 
$(\nu_\mu, \mu^-)$  reaction but overestimate the other two reactions. The HF+RPA  nuclear model of Auerbach et 
al.~\cite{Auerbach:1997ay} could  explain the $(\nu_e, e^-)$  and $(\nu_\mu, \mu^-)$ reactions but underestimate the ($\mu^-, 
\nu_\mu$) reactions. On the other hand, the elaborate shell model calculations with large basis space done by  Hayes and 
Towner~\cite{Hayes:1999ew} and Auerbach and Brown~\cite{Auerbach:2002tw} do better in explaining all the reactions qualitatively 
well but fail to reproduce quantitatively the cross section(capture rate) of $\nu_e$ and $\nu_\mu$ reactions ($\mu$-capture).

\item [(ii)]  In the case of Fermi gas models also the earlier non-relativistic calculations by Singh and Oset~\cite{Singh:1993rg}, 
Kosmas and Oset~\cite{Kosmas:1996fh}, 
and Nieves et al.~\cite{Nieves:2004wx, Valverde:2006zn} in the local Fermi gas model including RPA 
 reproduce ($\nu_e, e^-)$ but overestimate the ($\nu_\mu, \mu^-$) reaction cross section and 
underestimate $(\mu^-, \nu_\mu)$ capture rate. The 
local Fermi gas model has been considerably improved in the latest calculations by Nieves and Sobczyk~\cite{Nieves:2017lij},
Vagnoni et al.~\cite{Vagnoni:2017hll} and Ivanov et al.~\cite{Ivanov:2018nlm}, where relativistic transition operators and the spectral 
function of nucleons  are used to calculate all the three processes. These models reproduce ($\nu_e, e^-$) and ($\nu_\mu, \mu^-$) 
inclusive cross sections quite well but underestimate the inclusive ($\mu^-, \nu_\mu$) capture rates.
  \end{itemize}
  
It can be concluded from the results shown in Table-2 and the discussions given above that there exists real discrepancy between 
the theoretical and experimental results of the low energy weak inclusive processes of ($\nu_e, e^-$), ($\nu_\mu, \mu^-$) and 
($\mu^-, \nu_\mu$) in $^{12}$C. It should be noted that a discrepancy of about $15\%$ in the experimental cross section ($\nu_\mu, 
\mu^-$) is expected from the uncertainty due to the lack of the exact knowledge of the $\nu_\mu$ flux in the LSND experiment. On 
the theoretical side, it has been suggested that nuclear medium effects beyond the impulse approximation like the meson exchange 
currents and multinucleon effects  due to the ph-$\Delta$h and np-nh excitations may be important even in these low energy 
processes. However, some early calculation of these effects in the ($\nu_\mu, \mu^-$) reaction were not able to remove this 
discrepancy~\cite{Hayes:1999ew,Umino:1994wu,Umino:1996cz}. 
 
The important role of these nuclear medium effects beyond the impulse approximation was emphasized again almost a decade later 
when the theoretical attempts were made to understand the MiniBooNE results~\cite{AguilarArevalo:2010zc,Katori:2009du,
Sanchez:2009zz,Singh:2011zzp} on the inclusive muon neutrino scattering cross sections in ($\nu_\mu, \mu^-$) reaction in $^{12}$C  showed that 
their contributions is quite large and could explain the experimental results using the world average value of the axial dipole mass 
$M_A$ in the intermediate energy region of $E_{\nu_\mu}$=0.5 to 1GeV~\cite{Martini:2009uj, Martini:2010ex,Martini:2011wp,
Nieves:2011pp}. The impulse approximation calculations done to explain the experimentally observed cross sections 
required $M_{A}$ to be much higher than the world average value. This led to the MiniBooNE axial dipole mass ($M_A$) anomaly(section 
IIC). However, the calculations of the nuclear medium effects due to the multi-nucleon correlations and MEC effects in $\nu_\mu- 
^{12}$C scattering showed that their contribution is quite large and could explain the experimental results using the world 
average value of $M_A$. Such calculations are required to be done at the lower neutrino energies relevant to the LSND experiments 
with muon neutrinos. 
 
Therefore, in the case of LSND experiments, this  discrepancy could be a manifestation of the theoretical problem of not able to 
describe correctly enough the nuclear medium effect in $^{12}$C or an experimental problem of not doing a precise enough 
measurement of the inclusive cross-section specially in $(\nu_\mu-^{12}C)$ scattering in the energy region of muon neutrino from 
pion DIF.

\subsection{MiniBooNE Axial Dipole Mass~($M_{A}$) Anomaly and Nuclear Medium Effects}\label{sec3}
In the V-A theory of the charged weak interactions in the standard model, the inclusive neutrino scattering cross sections 
receive direct contribution from the vector currents and axial vector currents as well as from the interference between the 
vector and the axial vector currents. The direct contribution from the vector currents is expressed in terms of the weak vector 
form factors determined by the electric and magnetic Sachs' form factors $G_E(q^2)$ and 
$G_M(q^2)$~\cite{Punjabi:2015bba}, respectively. On the other hand, the direct contribution from the axial currents and the 
contribution from the interference term involve the axial vector form factor $g_A(q^2)$, which is parameterized as:
\begin{equation}
g_A(q^2)=\frac{g_A(0)}{\left(1-\frac{q^2}{M_A^2}\right)^2}
\end{equation}
where $g_A(0)$ is the axial charge determined from the neutron decay to be $g_A(0)=-1.26$ and $M_A$ is the axial dipole mass, generally 
taken to be the world average value of $M_A=1.026\pm0.016$GeV, determined from the quasielastic scattering or $M_A=1.014\pm0.016$ 
GeV determined from threshold pion electroproduction from proton/deuteron \cite{Bernard:2001rs, Bodek:2007ym}. However, using 
these values of $M_A$, the inclusive total cross sections and the differential cross sections, obtained from the high statistics 
experiment in $^{12}$C, performed by the MiniBooNE collaborations, were underestimated~\cite{AguilarArevalo:2010zc,Katori:2009du,
Sanchez:2009zz}. The MiniBooNE results were analyzed using the relativistic Fermi gas model and the microscopic nuclear models, 
which failed to explain the observed 
cross sections using the world average value of $M_A$~\cite{Singh:2011zzp}. It was also reported that a higher value of 
$M_A=1.35\pm 0.17$~\cite{Benhar:2006nr, Benhar:2013bwa} can explain both the total cross sections and the differential cross 
sections. This value of $M_A$ is considerably larger than the world value of $M_A$ determined from earlier experiments. The higher 
value of $M_A$ is also in disagreement with the results of another high statistics experiment recently performed by the NOMAD 
collaboration in $^{56}Fe$, which reported a value of $M_A=1.05\pm0.02\pm0.06$~GeV~ \cite{Lyubushkin:2008pe}, consistent with the 
world average value of $M_A$. This is known as the MiniBooNE axial dipole
mass anomaly. 

Assuming that the uncertainties in the neutrino flux at the MiniBooNE detector were well estimated and are reflected in the 
uncertainties quoted in the cross section measurements, there was a general consensus that the nuclear medium effects are not 
adequately taken into account. This may be because:
\begin{itemize}
\item[1.] The effect of nuclear medium beyond the impulse approximation like the meson exchange currents (MEC) and the 
$np-nh$ and the $ph-\Delta h$ excitations are not included adequately in these models, despite indications that they are quite 
important in the region of the low and intermediate energy neutrino reactions.
\item[2.] In the intermediate energy region of the MiniBooNE experiment, where the $\nu_\mu$ spectrum peaks around 750 MeV, the real pions 
would be produced which could be re-absorbed in the nuclear medium mimicking the genuine quasielastic inclusive events leading to an enhancement 
 in the observed inclusive cross sections for quasielastic reactions. The effect of these events called the quasielastic 
like(QE-like) events were not included adequately in the theoretical calculations.
\item[3.] In most of the neutrino reactions, the energy of the initial neutrinos is reconstructed using free particle quasielastic 
kinematics of neutrino-nucleon reactions in the nuclear medium. This kinematics is affected by the entanglement of the kinematics 
of the quasielastic-like events due to the inelastic processes i.e. $\nu_\mu N N\rightarrow \mu \Delta N \rightarrow \mu N N$ or 
scattering from the correlated pair $\nu_\mu N N \rightarrow  \mu N N$ in the nucleus with the genuine lepton events produced in 
the real quasielastic $\nu_\mu N\rightarrow \mu N$ scattering. The effect of this entanglement was not included in reconstructing 
the neutrino energy leading to underestimate the flux averaged cross sections.
\end{itemize}
A careful investigation of the above nuclear medium effects beyond impulse approximation was undertaken in view of the MiniBooNE 
axial dipole 
mass anomaly. The earlier calculations of Marteau et al. \cite{Marteau:1999kt}, Singh and Oset \cite{Singh:1993rg}, and
Nieves et al.~\cite{Nieves:2004wx, Valverde:2006zn} were, respectively, improved by Martini et al.~\cite{Martini:2010ex,
Martini:2011wp} and Nieves et al.~\cite{Nieves:2011pp} in which the $2p-2h$, $ph-\Delta h$, MEC effects as well as the pion 
reabsorption effects were taken 
into account. It was shown that the contribution of these effects is quite substantial in the energy region of MiniBooNE 
experiment and the observed cross sections are reproduced quite well when the above mentioned nuclear medium effects are taken 
into account using the world average value of $M_A$ and can explain the axial dipole mass anomaly. These results were further improved by the calculations of 
 Rocco et al.~\cite{Rocco:2018mwt} and Ivanov et al.~\cite{Ivanov:2018nlm} using the nucleon spectral functions to describe the momentum distribution 
 of the nucleons in the nucleus. 
\section{Inclusive Cross sections with monoenergetic KDAR neutrinos with $E_{\nu_\mu}$=236MeV}\label{sec4}
\subsection{Experimental and Theoretical results in the Impulse Approximation}
The monoenergetic muon neutrinos from KDAR  are identified as an ideal probe to study the neutrino nucleus 
cross sections in the low energy region in order to benchmark the nuclear medium effects in the exclusive as well as in the 
inclusive reactions in this energy region. An experiment for measuring neutrino cross section 
$\sigma(E_{\nu_\mu})$ with monoenergetic neutrinos would be free from the uncertainties arising from the reconstruction procedure 
of the initial neutrino energy present in most of the present experiments using beams of continuous energy of the muon neutrinos 
from accelerators. Some new experiments have been planned to measure inclusive cross sections in $^{12}$C and $^{40}$Ar using the 
monoenergetic neutrinos from kaons decaying at rest. The monoenergetic neutrinos from kaons decaying at rest are copiously 
produced with an energy $E_{\nu_{\mu}}=236$ MeV from the $K^+\rightarrow \mu^+ \nu_\mu$ decays along with a continuous energy 
spectrum of $\nu_e$ and $\nu_\mu$ from $Kl_3$ decays like $K^+ \rightarrow \pi^0 e^+ \nu_e$
 and 
$K^+\rightarrow \pi^0 \mu^+ \nu_\mu$.

The first measurements of the inclusive cross section in $^{12}$C nucleus with the monoenergetic KDAR muon neutrinos has been 
recently reported by the MiniBooNE collaboration to be $(2.7\pm0.9\pm0.8)\times 10^{-39}$ 
cm$^2$/neutron~\cite{Aguilar-Arevalo:2018ylq}. Adding the errors in quadrature, the collaboration has also quoted a value of 
$\sigma =(2.7\pm1.2)\times 10^{-39}cm^2$. Theoretically, this reaction in $^{12}$C in the low energy region of few hundreds of 
MeV, has been studied by many authors as discussed in section-\ref{sec2}, but specific calculations and discussions of the 
inclusive cross section for $E_{\nu_{\mu}}=236$ MeV have been done recently by Akbar et al. \cite{Akbar:2017dih} in the 
relativistic Fermi gas model with RPA to include the effect of correlations and Nikolakopoulos et al. ~\cite{Nikolakopoulos:2020alk} in 
a microscopic model using CRPA to include the effect of nucleon -nucleon correlations. Nikolakopoulos et al. \cite{Nikolakopoulos:2020alk} have also 
extrapolated the results of some earlier calculations to predict the inclusive neutrino cross sections at $E_{\nu_\mu}$=236MeV and 
presented a comparative study of the theoretical and experimental results. In Table-3, we present a list of the theoretical results 
for the inclusive cross sections at $E_{\nu_\mu}$=236MeV in the process $\nu_\mu +^{12}C\rightarrow \mu^{-}+X$ obtained in 
various theoretical calculations along with the experimental result from the MiniBooNE experiment~\cite{Aguilar-Arevalo:2018ylq} 
for comparison. 
\begin{table}
{\footnotesize
\begin{tabular}{ |c|c|c|c| c|}
 \hline
$\sigma(\nu_\mu)$~~~~~~~~~~~~&123.7$<E_{\nu_{\mu}}<$280MeV&159$\pm$26$\pm$37~\cite{Koetke:1992yk}~~~~~~~&8.3$\pm$0.7$\pm$1.6~\cite{Albert:1994xs}~~~~~~~~~&11.2$\pm$0.3$\pm$1.8~\cite{Athanassopoulos:1997rn}\\
$(\times 10^{-40}cm^2)$&&&12.4$\pm$0.3$\pm$1.8~\cite{Athanassopoulos:1997pv}&10.6$\pm$ 0.3$\pm$1.8~\cite{Auerbach:2002iy}~~~~~~\\\hline
$\sigma(\nu_e)$&$E_{\nu_{e}}<52.8$ MeV&14.1$\pm$2.3~\cite{Krakauer:1991rf}&15.2$\pm$1.0$\pm$1.3~\cite{Bodmann:1994py} &14.8$\pm$0.7 $\pm$1.1~
\cite{Athanassopoulos:1997rm}\\
$(\times 10^{-42}cm^2)$&&&&16.8$\pm$1.7~\cite{Kleinfelleret:1993}\\\hline
$\sigma(\nu_\mu)$&&2.7$\pm$0.9$\pm$0.8~\cite{Aguilar-Arevalo:2018ylq}&&\\
$(\times 10^{-39}cm^2/neutrons)$&$E_{\nu_{\mu}}=$ 236 MeV&&&\\\hline
$\Gamma(\mu^-)$&$m_\mu$=105.6 MeV &3.88 $\pm$ 0.05~\cite{Suzuki:1987jf}&&3.71$\pm$0.11~\cite{Eckhause:1966}\\
($10^4 s^{-1}$)&&&&3.77$\pm$0.07~\cite{Ishida:1986jz}\\
&&&&3.76$\pm$0.04~\cite{Martino:1982}\\\hline
 \end{tabular}
 }
 \caption{A summary of the experimental results for the flux averaged inclusive $\nu_e$ and $\nu_\mu$ scattering cross sections 
 and muon capture  rates in $^{12}$C.}
 \end{table}
 
\begin{table}
  \begin{tabular}{|c|c|c|c|}\hline
   Process & Experiments& Microscopic Theories& Fermi Gas  \\
   &&&Model\\\hline
   $\sigma(\nu_\mu)$&&28.8,22.4,14.5,15.2~\cite{Auerbach:1997ay}&16.65$\pm$1.37~\cite{Singh:1998md}\\
   $(\times 10^{-40})$&&27.0,21.1,13.5,14.3~\cite{Auerbach:1997ay}&19~\cite{Kosmas:1996fh}\\
   $cm^2$&12.4$\pm$0.3$\pm$1.8~\cite{Athanassopoulos:1997rn,Kolbe:1999au}&19.25~\cite{Kolbe:1995af},19.59~\cite{Paar:2008zza}&13.2$\pm$0.7,9.7$\pm$0.3,12.2~\cite{Nieves:2017lij}\\
   $\nu_{\mu} (^{12}C,X)\mu^{-}$&&18.18~\cite{Kolbe:1999au}&22.7-24.1~\cite{Umino:1996cz,Umino:1994wu}\\
   &&15.6,13.2,17.0,31.3,19.1~\cite{Hayes:1999ew} &25~\cite{Singh:1993rg}\\
   &&15.18,17.80,19.23,20.29,21.08~\cite{Volpe:2000zn}&22.7-24.1~\cite{Umino:1994wu,Umino:1996cz}\\
   &&30.0, 19.2~\cite{Auerbach:2002tw}&11.9~\cite{Nieves:2004wx}\\
\hline
   $\sigma(\nu_e)$ &&15.2,15.6~\cite{Kolbe:1994xb}&14~\cite{Nieves:2004wx},15.48$\pm$1.13~\cite{Singh:1998md}\\
   $(\times 10^{-42})$&& 16.42,16.70,55.1,52.0~\cite{Volpe:2000zn}&14~\cite{Kosmas:1996fh} \\
   $cm^2$&&19.28,~\cite{Kolbe:1999au}&13.8$\pm$0.4,14.3$\pm$0.6,8.6~\cite{Nieves:2017lij} \\ 
   &&6.9,3.5,4.1,5.4,3.1~\cite{Hayes:1999ew}& \\
   $\nu_{e} (^{12}C,X)e^{-}$& 14.8$\pm$0.7$\pm$1.4&23.7, 15.1~\cite{Auerbach:2002tw},15~\cite{Kolbe:2003ys},12.14~\cite{Paar:2008zza}& 15.3~\cite{Singh:1993rg} \\
   &&14.6~\cite{Donnelly:1978tz},13.1~\cite{Donnelly:1973enn}&13.6~\cite{SajjadAthar:2005ke}\\
   &&90.6,63.2,12.9,17.6~\cite{Auerbach:1997ay}&\\
  &&114.4,76.3,16.5,22.7~\cite{Auerbach:1997ay}& \\
  \hline
   $\Gamma(\mu^-)$ &3.88 $\pm$ 0.05~\cite{Suzuki:1987jf} &4.82,4.26,4.07,4.47~\cite{Krmpotic:2004gx} &3.3~\cite{Singh:1993rg}\\
   $(\times 10^4) $ &&&3.37$\pm$0.16,3.22,3.19$\pm$0.06~\cite{Nieves:2017lij}\\
  $sec^{-1}$ &&2.98,2.99,3.17,3.40~\cite{Hayes:1999ew}&3.60$\pm$0.22~\cite{Singh:1998md}\\
   $\mu^{-} (^{12}C,X)\nu_{\mu}$&&3.56,4.53~\cite{Hayes:1999ew}&\\
   && &\\
   &&&3.21~\cite{Nieves:2004wx}\\
   &&5.24,3.35~\cite{Auerbach:2002tw}&\\
   &&8.0,6.87,3.09,3.48~\cite{Auerbach:1997ay}&\\
   &&8.4,7.22,3.23,3.64~\cite{Auerbach:1997ay}&\\
  &&3.32,4.06,5.12,5.79~\cite{Volpe:2000zn}& \\
  &&&\\
  \hline
   
  \end{tabular}
\caption{Latest experimental results and various theoretical results in different nuclear models for inclusive cross section for 
$\nu_e$ and $\nu_\mu $ scattering and muon capture rates in$^{12}$C.}
 \end{table}
\begin{table}
 \begin{tabular}{|c|c|}\hline
  Experimental and Theoretical Models~~~~~~~~~~~~~~~~& Cross section\\\hline
  MiniBooNE Exp.~\cite{Aguilar-Arevalo:2018ylq}  & 2.7$\pm$1.2\\\hline
  Akbar et al.~\cite{Akbar:2017dih} & 0.91\\\hline
  Martini et al. ~\cite{Martini:2011wp,Auerbach:2001hz}& 1.3+0.2(np-nh)\\\hline
  NuWro~\cite{Juszczak:2009qa,Golan:2012wx} & 1.3+0.4(np-nh)\\\hline
  GENIE~\cite{Andreopoulos:2009rq} & 1.75\\\hline
  NUANCE~\cite{Casper:2002sd} & 1.4\\
  \hline
  CRPA~\cite{Gonzalez-Jimenez:2019qhq} & 1.58\\\hline
  RMF~\cite{Gonzalez-Jimenez:2019qhq} & 1.56\\\hline
  RFG~\cite{Nikolakopoulos:2020alk} &1.66\\\hline
  RFG 34~\cite{Nikolakopoulos:2020alk} &1.38\\\hline
 \end{tabular}
\caption{Experimental and theoretical results for the inclusive cross section for KDAR neutrinos. The cross sections are in units of $10^{-39}cm^2$.}
\end{table}

We see that all the theoretical predictions for the theoretical cross sections lie in the wide range of $(0.91 ~\text{to}~ 1.66)\times 10^{-39}cm^2/neutron$ and 
underestimate the experimental results for the inclusive cross sections at $E_{\nu_\mu}$=236MeV. A comparison of the theoretical 
results of the 
inclusive neutrino cross sections in $^{12}$C in the case of monoenergetic neutrinos at $E_{\nu_\mu}$=236MeV and the 
earlier theoretical results in the energy region of the LSND experiments i.e. $E_{\nu_\mu} <$ 280MeV with 
the 
experimental data, discussed in section-2, shows that :
\begin{itemize}
\item[(i)] The theoretical predictions using various nuclear models for the inclusive cross sections for the reaction $\nu_\mu 
+^{12}C\rightarrow \mu^- +X$ have a large range of variation. This is surprising in the case of $^{12}$C which is one of the 
 theoretically better studied nucleus.
\item[(ii)] Most of the theoretical predictions for the inclusive cross section for this reaction overestimate the experimental results for the LSND experiment with neutrino energy $E_{\nu_{\mu}}$ in the range of $120 MeV < E_{\nu_{\mu}} < 280$ MeV, while the theoretical predictions in the case of the KDAR neutrino with $E_{\nu_{\mu}}=236$ MeV underestimate the experimental result. 
\item[(iii)] The latest experimental as well as theoretical results for the inclusive cross sections using the monoenergetic KDAR muon 
neutrinos together with the results obtained in the case of LSND and KARMEN experiments with electron and muon neutrinos along with the capture rate of 
$(\mu^-, \nu_\mu)$ process in $^{12}$C show that the theoretical calculations done in the 
impulse approximation for all the weak nuclear processes of $(\mu^- , \nu_\mu), ~(\nu_e , e^-)$ and $(\nu_\mu , \mu^- )$ in the
low energy region are not theoretically understood satisfactorily, with a given nuclear model used to describe the structure of $^{12}$C nucleus.
\end{itemize}
\subsection{Beyond the Impulse Approximation}
We have discussed the importance of the nuclear medium effects like the 2p-2h, ph-$\Delta$h and meson exchange currents beyond the 
impulse approximation in the case of inclusive neutrino scattering in $^{12}$C in the intermediate energy region of several 
hundreds of MeV. These effects were also shown earlier to be important in the very low energy region of 
the nuclear beta decays~\cite{Dautry:1975xq,Ohta:1974fa,Ericson:1973vj,Riska:1970jxh, Nieves:2011yp,Delorme:1985ps,Muller:1981dc},
solar neutrino reaction~\cite{Gari} 
and muon capture~\cite{Suzuki:1987jf, Mukhopadhyay:1998me}. However, very few studies 
of these effects have been made in the case of inclusive neutrino reactions in $^{12}$C relevant to the neutrino energies of 
$\pi$-DIF and KDAR neutrinos, for example, the early works of Hayes and Towner~\cite{Hayes:1999ew} 
 and Umino et al.~\cite{Umino:1994wu,Umino:1996cz}. 
 in the case of $\pi$-DIF neutrinos and the recent work of Nikolakopoulos et al.~\cite{Nikolakopoulos:2020alk} in the case of 
 KDAR neutrinos.
 
 In the case of $\pi$-DIF neutrinos, the calculations of Hayes and Towner~\cite{Hayes:1999ew} were performed in a microscopic nuclear model
 using a multiparticle shell model with large basis space, while the calculations of Umino et al.~\cite{Umino:1994wu,Umino:1996cz} were done 
 using the relativistic Fermi gas model. Both the calculations find a reduction of about 20$\%$ in the inclusive cross sections while in the
 case of KDAR neutrinos, the work of Nikolakopoulos et al.~\cite{Nikolakopoulos:2020alk} finds an increase of about 20-25$\%$ obtained from extrapolating 
 the works of NuWro~\cite{Juszczak:2009qa, Golan:2012wx} and Martini et al.~\cite{Martini:2010ex, Martini:2011wp}. 
 
 Recently an ab initio calculation of the inclusive $\nu_\mu$ cross section in $^{12}$C has been done by Rocco et 
 al.~\cite{Rocco:2018mwt}
 including the contribution of some two body effects. Moreover, Ivanov et al.~\cite{Ivanov:2018nlm} have made an improvement over the 
 calculations 
 of 
 Nieves et al.~\cite{Nieves:2011pp} by using an spectral function $S(p,E)$ for the nucleon momentum distribution to calculate the 
 inclusive cross sections in the relativistic Fermi gas model.
  We show in 
 Fig.3, the results for the inclusive cross section $\sigma(E_{\nu_\mu})$ as a function of the neutrino energy E$_{\nu_\mu}$ in
 the energy range $0<E_{\nu_\mu}<500$MeV in various models. It is clear from Fig.3 that the calculations
 by Rocco et al.~\cite{Rocco:2018mwt} and Ivanov et al.~\cite{Ivanov:2018nlm}
  show an enhancement in the 
 inclusive cross section at $E_{\nu_\mu}$=236MeV which are quantitatively small as compared to the results quoted by 
 Nikolakopoulos et al.~\cite{Nikolakopoulos:2020alk}.
 
 
We observe from the results shown in Table-3, that:
\begin{itemize}
\item[(i)] The contribution of the nuclear medium effects beyond the impulse approximation  is to increase the inclusive cross
 sections but the increase is not sufficient enough to explain the results of the KDAR neutrinos.
\item[(ii)] Such an increase in the inclusive cross section in the theoretical predictions due to the nuclear medium effects, in 
the similar  
energy region of the $\pi$-DIF neutrinos would further enhance the disagreement between the theoretical and the experimental results in the case of LSND experiment. 
\item[(iii)] Moreover, this would also be in contradiction with earlier results of such effects calculated in the work of Hayes and Towner~\cite{Hayes:1999ew} in microscopic models and Umino et al.~\cite{Umino:1994wu,Umino:1996cz} in the case of Fermi gas models.
\end{itemize}
It is clear that present status of the theoretical calculations for the inclusive cross section in the process $\nu_\mu +^{12}C\rightarrow\mu^- +X$ in the low energy region of few hundreds of MeV is not satisfactory even with the inclusion of nuclear medium effects beyond the impulse approximation calculated in various models available in the literature.

\section{Summary and Conclusions}\label{Summary}
The experimental and the theoretical status of the low energy inclusive neutrino scattering cross sections and inclusive 
muon capture from $^{12}$C nucleus has been reviewed in the context of new measurements reported by the MiniBooNE collaboration with monoenergetic 
muon neutrinos from KDAR with energy $E_{\nu_\mu}=236$ MeV. To summarize, we find that
\begin{itemize}
\item[(1)] The various theoretical predictions for the inclusive cross sections in the case of KDAR neutrinos underestimate the experimentally
observed cross sections showing a discrepancy between the theory and experiment. This reminds us of the discrepancy observed between the experimental
and theoretical results in the case of the low energy measurements of the inclusive cross section reported about 20 years ago by the LSND 
experiments using muon neutrinos from $\pi$DIF, in the energy range of $120<E_{\nu_\mu}<280$ MeV, with the 
difference 
that in the case of LSND experiment the results were overestimated by the theoretical predictions.
\item[(2)]  There is no theoretical model which succeeds in reproducing the inclusive scattering cross sections and the inclusive capture rates, 
respectively, in all the three weak nuclear processes at low energies induced by the muon neutrinos, electron neutrinos and muons in $^{12}$C.
This is indeed surprising in the case of $^{12}$C nucleus, which is considered to be one of the best known nucleus regarding its structure. 
This calls for a better understanding of the nuclear medium effects in the low energy weak interaction processes induced by neutrinos and 
muons in $^{12}$C as well as in other nuclei.
\item[(3)] The recent theoretical calculations of the inclusive neutrino cross sections in $^{12}$C in the region of intermediate energy of neutrinos
around 1 GeV show that the role of meson exchange currents, $2p-2h$ and $\Delta h$ excitations etc., is quite important in explaining the inclusive 
neutrino cross sections in $^{12}$C observed in the MiniBooNE experiment to explain the axial dipole mass anomaly. These calculations when extended to the 
low energy region of neutrinos from pion decays in flight corresponding to the LSND experiments and the neutrinos from the kaon decays
at rest, do not help to remove the discrepancy between the theoretical and experimental results.
\item[(4)] In view of the LSND experiments being completed almost 20 years ago, it is desirable that future experiments planned at the Fermilab and JPARC laboratories
 with
the low energy monoenergetic KDAR neutrinos are performed with enhanced precision and the theoretical calculations are done using the advanced state 
of the art methods to compute the nuclear medium effects including the many body effects for the low energy inclusive reactions in $^{12}$C and $^{40}$Ar. 
\end{itemize}


\begin{thebibliography}{}
       \bibitem{Aguilar-Arevalo:2018ylq} 
  A.~A.~Aguilar-Arevalo {\it et al.} [MiniBooNE Collaboration],
  Phys.\ Rev.\ Lett.\  {\bf 120}, 141802 (2018).
  
  \bibitem{Akbar:2017dih} 
  F.~Akbar, M.~Sajjad Athar and S.~K.~Singh,
  J.\ Phys.\ G {\bf 44}, 125108 (2017).
   \bibitem{Nikolakopoulos:2020alk} 
  A.~Nikolakopoulos, V.~Pandey, J.~Spitz and N.~Jachowicz,
  arXiv:2010.05794 [nucl-th].

\bibitem{Cortex} B. Cortex et al., Proceedings of 1981 Orbis Scientific Conference, Edited by A. Perlmutter.
\bibitem{Koetke:1992yk} 
  D.~D.~Koetke {\it et al.},
  Phys.\ Rev.\ C {\bf 46}, 2554 (1992).
 
  \bibitem{Albert:1994xs} 
  M.~Albert {\it et al.} [LSND Collaboration],
  Phys.\ Rev.\ C {\bf 51}, 1065 (1995).
  
  \bibitem{Athanassopoulos:1997rn} 
  C.~Athanassopoulos {\it et al.} [LSND Collaboration],
  Phys.\ Rev.\ C {\bf 56}, 2806 (1997).
  
  \bibitem{Auerbach:2002iy} 
  L.~B.~Auerbach {\it et al.} [LSND Collaboration],
  Phys.\ Rev.\ C {\bf 66}, 015501 (2002).
  
 
  
  \bibitem{Krakauer:1991rf} 
  D.~A.~Krakauer {\it et al.},
  Phys.\ Rev.\ C {\bf 45}, 2450 (1992).
 
  \bibitem{Bodmann:1994py} 
  B.~E.~Bodmann {\it et al.} [KARMEN Collaboration],
  Phys.\ Lett.\ B {\bf 332}, 251 (1994).
  \bibitem{Kretschmer:2002iq} 
  W.~Kretschmer [KARMEN Collaboration],
  Acta Phys.\ Polon.\ B {\bf 33}, 1775 (2002).
  \bibitem{Maschuw:1998qh} 
  R.~Maschuw [KARMEN Collaboration],
  Prog.\ Part.\ Nucl.\ Phys.\  {\bf 40}, 183 (1998).
  \bibitem{Athanassopoulos:1997rm} 
  C.~Athanassopoulos {\it et al.} [LSND Collaboration],
  Phys.\ Rev.\ C {\bf 55}, 2078 (1997).
  

  
  \bibitem{Auerbach:2001hz} 
  L.~B.~Auerbach {\it et al.} [LSND Collaboration],
  Phys.\ Rev.\ C {\bf 64}, 065501 (2001).
  
  \bibitem{LlewellynSmith:1971uhs} 
  C.~H.~Llewellyn Smith,
  Phys.\ Rept.\  {\bf 3}, 261 (1972).
  
    \bibitem{Walecka} J. D. Walecka in Muon Physics Vol.2, Edited by V. W. Hughes and C. S. Wu, Academic Press (1975)

  
%
  \bibitem{Donnelly:1978tz} 
  T.~W.~Donnelly and R.~D.~Peccei,
  Phys.\ Rept.\  {\bf 50}, 1 (1979).
  

  
  \bibitem{Kolbe:1994xb} 
  E.~Kolbe, K.~Langanke and S.~Krewald,
  Phys.\ Rev.\ C {\bf 49}, 1122 (1994).
  
  \bibitem{Kolbe:1995af} 
  E.~Kolbe, F.~K.~Thielemann, K.~Langanke and P.~Vogel,
  Phys.\ Rev.\ C {\bf 52}, 3437 (1995).
  
  \bibitem{Kolbe:1999au} 
  E.~Kolbe, K.~Langanke and P.~Vogel,
  Nucl.\ Phys.\ A {\bf 652}, 91 (1999).
  
  \bibitem{Kolbe:2003ys} 
  E.~Kolbe, K.~Langanke, G.~Martinez-Pinedo and P.~Vogel,
  J.\ Phys.\ G {\bf 29}, 2569 (2003).
 
  \bibitem{Volpe:2000zn} 
  C.~Volpe, N.~Auerbach, G.~Colo, T.~Suzuki and N.~Van Giai,
  Phys.\ Rev.\ C {\bf 62}, 015501 (2000).
  
  \bibitem{Auerbach:1997ay} 
  N.~Auerbach, N.~Van Giai and O.~K.~Vorov,
  Phys.\ Rev.\ C {\bf 56}, R2368 (1997).
  
  \bibitem{Hayes:1999ew} 
  A.~C.~Hayes and I.~S.~Towner,
  Phys.\ Rev.\ C {\bf 61}, 044603 (2000).
  
  \bibitem{Auerbach:2002tw} 
  N.~Auerbach and B.~A.~Brown,
  Phys.\ Rev.\ C {\bf 65}, 024322 (2002).
  
  \bibitem{Paar:2008zza} 
  N.~Paar, D.~Vretenar and P.~Ring,
  J.\ Phys.\ G {\bf 35}, 014058 (2008).
  
  \bibitem{Krmpotic:2004gx} 
  F.~Krmpotic, A.~Samana and A.~Mariano,
  Phys.\ Rev.\ C {\bf 71}, 044319 (2005).
  
  \bibitem{Pandey:2014tza} 
  V.~Pandey, N.~Jachowicz, T.~Van Cuyck, J.~Ryckebusch and M.~Martini,
  Phys.\ Rev.\ C {\bf 92}, 024606 (2015).
  
  
  \bibitem{Jachowicz:2002rr} 
  N.~Jachowicz, K.~Heyde, J.~Ryckebusch and S.~Rombouts,
  Phys.\ Rev.\ C {\bf 65}, 025501 (2002).
  
  \bibitem{Gonzalez-Jimenez:2019qhq} 
  R.~González-Jiménez, A.~Nikolakopoulos, N.~Jachowicz and J.~M.~Udías,
  Phys.\ Rev.\ C {\bf 100}, 045501 (2019).
%
  
    \bibitem{Smith:1972xh} 
  R.~A.~Smith and E.~J.~Moniz,
  Nucl.\ Phys.\ B {\bf 43}, 605 (1972).
  Erratum: [Nucl.\ Phys.\ B {\bf 101}, 547 (1975)].
  
  \bibitem{Gaisser:1986bv} 
  T.~K.~Gaisser and J.~S.~O'Connell,
  Phys.\ Rev.\ D {\bf 34}, 822 (1986).
  
  \bibitem{Singh:1993rg} 
  S.~K.~Singh and E.~Oset,
  Phys.\ Rev.\ C {\bf 48}, 1246 (1993).
    \bibitem{Umino:1994wu} 
  Y.~Umino, J.~M.~Udias and P.~J.~Mulders,
  Phys.\ Rev.\ Lett.\  {\bf 74}, 4993 (1995).
  \bibitem{Umino:1996cz} 
  Y.~Umino and J.~M.~Udias,
  Phys.\ Rev.\ C {\bf 52}, 3399 (1995).
  \bibitem{Kosmas:1996fh} 
  T.~S.~Kosmas and E.~Oset,
  Phys.\ Rev.\ C {\bf 53}, 1409 (1996).
 
  \bibitem{Singh:1998md} 
  S.~K.~Singh, N.~C.~Mukhopadhyay and E.~Oset,
  Phys.\ Rev.\ C {\bf 57}, 2687 (1998).
  
  \bibitem{Nieves:2004wx} 
  J.~Nieves, J.~E.~Amaro and M.~Valverde,
  Phys.\ Rev.\ C {\bf 70}, 055503 (2004)
  Erratum: [Phys.\ Rev.\ C {\bf 72}, 019902 (2005)].
  
    \bibitem{SajjadAthar:2005ke} 
  M.~Sajjad Athar, S.~Ahmad and S.~K.~Singh,
  Nucl.\ Phys.\ A {\bf 764}, 551 (2006).

 
  \bibitem{Valverde:2006zn} 
  M.~Valverde, J.~E.~Amaro and J.~Nieves,
  Phys.\ Lett.\ B {\bf 638}, 325 (2006).
  
    \bibitem{Nieves:2017lij} 
  J.~Nieves and J.~E.~Sobczyk,
  Annals Phys.\  {\bf 383}, 455 (2017).
   
  \bibitem{Vagnoni:2017hll} 
  E.~Vagnoni, O.~Benhar and D.~Meloni,
  Phys.\ Rev.\ Lett.\  {\bf 118}, 142502 (2017).
 

   
  
   \bibitem{Ivanov:2018nlm} 
  M.~V.~Ivanov {\it et al.},
  Phys.\ Rev.\ C {\bf 99}, 014610 (2019)
  
  \bibitem{Mintz:1995ww} 
  S.~L.~Mintz and M.~Pourkaviani,
  Nucl.\ Phys.\ A {\bf 594}, 346 (1995).
  
  \bibitem{Kim:1985zs} 
  C.~W.~Kim and S.~L.~Mintz,
  Phys.\ Rev.\ C {\bf 31}, 274 (1985).
  \bibitem{Frazier:1970rb} 
  J.~Frazier, C.~W.~Kim and M.~Ram,
  Phys.\ Rev.\ D {\bf 1}, 3168 (1970)
  Erratum: [Phys.\ Rev.\ D {\bf 3}, 271 (1971)].
  
  \bibitem{Kim:1979}
  C. W. Kim and H. Primakoff, Mesons in Nuclei, Vol. 1, Ed. M. Rho and D. Wilkinson, 67 (1979) 
  
  \bibitem{Kubodera:1993rk} 
  K.~Kubodera and S.~Nozawa,
  Int.\ J.\ Mod.\ Phys.\ E {\bf 3}, 101 (1994).
 
  \bibitem{Alvarez-Ruso:2017oui} 
  L.~Alvarez-Ruso {\it et al.} [NuSTEC Collaboration],
  Prog.\ Part.\ Nucl.\ Phys.\  {\bf 100}, 1 (2018).
  
  \bibitem{Alvarez-Ruso:2014bla} 
  L.~Alvarez-Ruso, Y.~Hayato and J.~Nieves,
  New J.\ Phys.\  {\bf 16}, 075015 (2014).
  
  \bibitem{Morfin:2012kn} 
  J.~G.~Morfin, J.~Nieves and J.~T.~Sobczyk,
  Adv.\ High Energy Phys.\  {\bf 2012}, 934597 (2012).
  
  \bibitem{Formaggio:2013kya} 
  J.~A.~Formaggio and G.~P.~Zeller,
  Rev.\ Mod.\ Phys.\  {\bf 84}, 1307 (2012).
  
  \bibitem{Gallagher:2011zza} 
  H.~Gallagher, G.~Garvey and G.~P.~Zeller,
  Ann.\ Rev.\ Nucl.\ Part.\ Sci.\  {\bf 61}, 355 (2011).
 
  \bibitem{Katori:2016yel} 
  T.~Katori and M.~Martini,
  J.\ Phys.\ G {\bf 45}, 013001 (2018).
   \bibitem{Benhar:2013bwa} 
  O.~Benhar and N.~Rocco,
  Adv.\ High Energy Phys.\  {\bf 2013}, 912702 (2013).
  \bibitem{Benhar:2015wva} 
  O.~Benhar, P.~Huber, C.~Mariani and D.~Meloni,
  Phys.\ Rept.\  {\bf 700}, 1 (2017).
    
    \bibitem{Spitz:2012gp} 
  J.~Spitz,
  Phys.\ Rev.\ D {\bf 85}, 093020 (2012).
  
  \bibitem{Spitz:2014hwa} 
  J.~Spitz,
  Phys.\ Rev.\ D {\bf 89}, 073007 (2014).
  
  \bibitem{Axani:2015dha} 
  S.~Axani, G.~Collin, J.~Conrad, M.~Shaevitz, J.~Spitz and T.~Wongjirad,
  Phys.\ Rev.\ D {\bf 92}, 092010 (2015).
  
   \bibitem{Acciarri:2016smi} 
  R.~Acciarri {\it et al.} [MicroBooNE Collaboration],
  JINST {\bf 12}, P02017 (2017).
  
  \bibitem{Antonello:2015lea} 
  M.~Antonello {\it et al.} [MicroBooNE and LAr1-ND and ICARUS-WA104 Collaborations],
  arXiv:1503.01520 [physics.ins-det].
  
     \bibitem{Fava:2019fuz} 
  A.~Fava [SBN Collaboration],
  PoS NuFACT {\bf 2018}, 011 (2019).
  
  \bibitem{Ajimura:2017fld} 
  S.~Ajimura {\it et al.},
  arXiv:1705.08629 [physics.ins-det].
  
  \bibitem{Harada:2013yaa} 
  M.~Harada {\it et al.} [JSNS2 Collaboration],
  arXiv:1310.1437 [physics.ins-det].
  
  \bibitem{Park:2020uck} 
  J.~S.~Park {\it et al.},
  JINST {\bf 15}, T07003 (2020).
  
  \bibitem{Park:2020vxw} 
  J.~S.~Park {\it et al.},
  New Phys.\ Sae Mulli {\bf 70}, 928 (2020).
    
    
  
  \bibitem{Mukhopadhyay:1976hu} 
  N.~C.~Mukhopadhyay,
  Phys.\ Rept.\  {\bf 30}, 1 (1977).


  
  \bibitem{Measday:2001yr} 
  D.~F.~Measday,
  Phys.\ Rept.\  {\bf 354}, 243 (2001).
  
   \bibitem{Auerbach:1984ph} 
  N.~Auerbach and A.~Klein,
  Nucl.\ Phys.\ A {\bf 422}, 480 (1984).
  
  
  
  \bibitem{Suzuki:1987jf} 
  T.~Suzuki, D.~F.~Measday and J.~P.~Roalsvig,
  Phys.\ Rev.\ C {\bf 35}, 2212 (1987).
 
 

\bibitem{Dautry:1975xq} 
  F.~Dautry, M.~Rho and D.~O.~Riska,
  Nucl.\ Phys.\ A {\bf 264}, 507 (1976).
  
  \bibitem{Ohta:1974fa} 
  K.~Ohta and M.~Wakamatsu,
  Phys.\ Lett.\  {\bf 51B}, 325 (1974).
  
  \bibitem{Ericson:1973vj} 
  M.~Ericson, A.~Figureau and C.~Thevenet,
  Phys.\ Lett.\  {\bf 45B}, 19 (1973).
  
  \bibitem{Riska:1970jxh} 
  D.~O.~Riska and G.~E.~Brown,
  Phys.\ Lett.\  {\bf 32B}, 662 (1970).
  
  
  
  \bibitem{Delorme:1985ps} 
  J.~Delorme and M.~Ericson,
  Phys.\ Lett.\  {\bf 156B}, 263 (1985).
  
    \bibitem{Nieves:2011yp} 
  J.~Nieves, I.~Ruiz Simo and M.~J.~Vicente Vacas,
  Phys.\ Lett.\ B {\bf 707}, 72 (2012).
  
  \bibitem{Muller:1981dc} 
  W.~Muller and M.~Gari,
  Phys.\ Lett.\  {\bf 102B}, 389 (1981).

    \bibitem{Divari:2012cj} 
  P.~C.~Divari, S.~Galanopoulos and G.~A.~Souliotis,
  J.\ Phys.\ G {\bf 39}, 095204 (2012).
   \bibitem{Totani:1997vj} 
  T.~Totani, K.~Sato, H.~E.~Dalhed and J.~R.~Wilson,
  Astrophys.\ J.\  {\bf 496}, 216 (1998).
  
  \bibitem{Duan:2010bf} 
  H.~Duan and A.~Friedland,
  Phys.\ Rev.\ Lett.\  {\bf 106}, 091101 (2011).
  
  \bibitem{Gava:2009pj} 
  J.~Gava, J.~Kneller, C.~Volpe and G.~C.~McLaughlin,
  Phys.\ Rev.\ Lett.\  {\bf 103}, 071101 (2009).

\bibitem{Scholberg:2012id} 
  K.~Scholberg,
  Ann.\ Rev.\ Nucl.\ Part.\ Sci.\  {\bf 62}, 81 (2012).
  
  \bibitem{Martino:1986gq} 
  J.~Martino,
  Nucl.\ Phys.\ A {\bf 453}, 591 (1986); J.~Martino,
  Czech.\ J.\ Phys.\ B {\bf 36}, 368 (1986); P.~Kammel {\it et al.},
  Nucl.\ Phys.\ A {\bf 663}, 911 (2000).
  
     
  
  \bibitem{Eckhause:1966}
  M.  Eckhause,  R.  T.  Siegel,  R.  E.  Welsh,  and  T.  A.  Filippas, Nucl. Phys. {\bf 81}, 575 (1966).
  
  
\bibitem{Ishida:1986jz} 
  K.~Ishida, J.~H.~Brewer, T.~Matsuzaki, Y.~Kuno, J.~Imazato and K.~Nagamine,
  Phys.\ Lett.\  {\bf 167B}, 31 (1986).
  
  \bibitem{Donnelly:1973enn} 
  T.~W.~Donnelly,
  Phys.\ Lett.\  {\bf 43B}, 93 (1973).
  
  
  
  \bibitem{AguilarArevalo:2010zc} 
  A.~A.~Aguilar-Arevalo {\it et al.} [MiniBooNE Collaboration],
  Phys.\ Rev.\ D {\bf 81}, 092005 (2010).
  
  \bibitem{Katori:2009du} 
  T.~Katori [MiniBooNE Collaboration],
  AIP Conf.\ Proc.\  {\bf 1189}, 139 (2009)
  
  \bibitem{Martini:2009uj} 
  M.~Martini, M.~Ericson, G.~Chanfray and J.~Marteau,
  Phys.\ Rev.\ C {\bf 80}, 065501 (2009).

  
  \bibitem{Martini:2010ex} 
  M.~Martini, M.~Ericson, G.~Chanfray and J.~Marteau,
  Phys.\ Rev.\ C {\bf 81}, 045502 (2010).
  
  \bibitem{Martini:2011wp} 
  M.~Martini, M.~Ericson and G.~Chanfray,
  Phys.\ Rev.\ C {\bf 84}, 055502 (2011).
  
  \bibitem{Nieves:2011pp} 
  J.~Nieves, I.~Ruiz Simo and M.~J.~Vicente Vacas,
  Phys.\ Rev.\ C {\bf 83}, 045501 (2011).
  
    
  \bibitem{Nieves:2012yz} 
  J.~Nieves, F.~Sanchez, I.~Ruiz Simo and M.~J.~Vicente Vacas,
  Phys.\ Rev.\ D {\bf 85}, 113008 (2012).
  \bibitem{Rocco:2018mwt} 
  N.~Rocco, C.~Barbieri, O.~Benhar, A.~De Pace and A.~Lovato,
  Phys.\ Rev.\ C {\bf 99}, 025502 (2019).
  
    \bibitem{Sanchez:2009zz} 
  F.~Sanchez, M.~Sorel and L.~Alvarez-Ruso,
  AIP Conf.\ Proc.\  {\bf 1189}, pp.1-358 (2009).
  
      \bibitem{Singh:2011zzp} 
  S.~K.~Singh, J.~G.~Morfin, M.~Sakuda and K.~D.~Purohit,
  AIP Conf.\ Proc.\  {\bf 1405}, pp.1 (2011).


  
  
\bibitem{Punjabi:2015bba} 
  V.~Punjabi, C.~F.~Perdrisat, M.~K.~Jones, E.~J.~Brash and C.~E.~Carlson,
  Eur.\ Phys.\ J.\ A {\bf 51}, 79 (2015).
  

    
  \bibitem{Bernard:2001rs} 
  V.~Bernard, L.~Elouadrhiri and U.~G.~Meissner,
  J.\ Phys.\ G {\bf 28}, R1 (2002).
  \bibitem{Bodek:2007ym} 
  A.~Bodek, S.~Avvakumov, R.~Bradford and H.~S.~Budd,
  Eur.\ Phys.\ J.\ C {\bf 53}, 349 (2008).
  
\bibitem{Benhar:2006nr}
O.~Benhar and D.~Meloni,
Nucl. Phys. A \textbf{789}, 379-402 (2007).
%
     \bibitem{Lyubushkin:2008pe} 
  V.~Lyubushkin {\it et al.} [NOMAD Collaboration],
  Eur.\ Phys.\ J.\ C {\bf 63}, 355 (2009).
  
    \bibitem{Marteau:1999kt} 
  J.~Marteau,
  Eur.\ Phys.\ J.\ A {\bf 5}, 183 (1999).
  
  \bibitem{Athanassopoulos:1997pv} 
  C.~Athanassopoulos {\it et al.} [LSND Collaboration],
  Phys.\ Rev.\ Lett.\  {\bf 81}, 1774 (1998).
%
  \bibitem{Kleinfelleret:1993}
  J. Kleinfellner {\it et al.}, KARMEN collaboration, in Proceedings of the XIII International Conference on Particles and Nuclei, Perugia, Italy, 1993, edited by A. Pascolini.

    \bibitem{Martino:1982}
J. Martino, Th`ese d’Etat, Uni. Paris-Sud (1982).

\bibitem{Juszczak:2009qa}
C.~Juszczak,
Acta Phys. Po. B {\bf 40}, 2507 (2009).

\bibitem{Golan:2012wx}
T.~Golan, C.~Juszczak and J.~T.~Sobczyk,
Phys. Rev. C \textbf{86}, 015505 (2012).

\bibitem{Andreopoulos:2009rq}
C.~Andreopoulos, A.~Bell, D.~Bhattacharya, F.~Cavanna, J.~Dobson, S.~Dytman, H.~Gallagher, P.~Guzowski, R.~Hatcher and P.~Kehayias, \textit{et al.}
Nucl. Instrum. Meth. A \textbf{614}, 87-104 (2010).

\bibitem{Casper:2002sd}
D.~Casper,
Nucl. Phys. B Proc. Suppl. \textbf{112}, 161-170 (2002).

\bibitem{Gari}
M. Gari and A. H. Hufmann,
APJ letters {\bf 174} (1972) L151; APJ {\bf 178} (1972) 543.
  \bibitem{Mukhopadhyay:1998me} 
  N.~C.~Mukhopadhyay, H.~C.~Chiang, S.~K.~Singh and E.~Oset,
  Phys.\ Lett.\ B {\bf 434}, 7 (1998).



\end{thebibliography}
\end{document}